\documentclass[submission, Phys]{SciPost}
\pdfoutput=1
\usepackage{comment}
\usepackage{enumerate}
\usepackage{array}
\usepackage{amssymb}
\usepackage{amsmath}
\usepackage{braket}
\usepackage{graphicx}
\usepackage{dsfont}
\usepackage{booktabs}
\usepackage{empheq}
\usepackage{braket}

\newcommand{\be}{\begin{equation}}
\newcommand{\ee}{\end{equation}}
\newcommand{\ba}{\begin{aligned}}
\newcommand{\ea}{\end{aligned}}
\newcommand{\bw}{\begin{widetext}}
\newcommand{\ew}{\end{widetext}}

\newcommand{\bea}{\begin{eqnarray}}
\newcommand{\eea}{\end{eqnarray}}

\begin{document}

\begin{center}{\Large \textbf{
Out-of-equilibrium   inhomogeneous XX  chains: Exact results and  the hydrodynamic limit 
}}\end{center}

% TODO: write the author list here. Use initials + surname format.
% Separate subsequent authors by a comma, omit comma at the end of the list.
% Mark the corresponding author with a superscript *.
\begin{center}
Vincenzo Alba\textsuperscript{1*}
Federico Rottoli\textsuperscript{1, 2}
\end{center}

\begin{center}
{\bf 1}	
Dipartimento di Fisica dell' Universit\`a di Pisa and INFN, Sezione di Pisa, I-56127 Pisa, Italy\\
{\bf 2}	
Istituto di Informatica e Telematica, CNR, I-56127 Pisa, Italy\\

* vincenzo.alba@unipi.it
\end{center}

\begin{center}
\today
\end{center}

% For convenience during refereeing: line numbers
%\linenumbers

\section*{Abstract}
{\bf 

We study the out-of-equilibrium dynamics in the XX chain with step-like magnetic field, which maps to 
an inhomogeneous tight-binding chain after Jordan-Wigner transformation. 
We obtain exact analytic expressions for the fermionic two-point correlation function  
after a quantum quench from several initial product states, both homogeneous and inhomogeneous ones. 
This is achieved by using a combination of Fourier and Laplace transforms, which allow us to map the problem 
to a standard Riemann-Hilbert problem on the unit circle. For arbitrary positions and times the correlators 
are not expressed in terms of elementary functions. However, in the hydrodynamic limit $x,y,t\to\infty$ with fixed
ratios, we provide explicit formulas that depend only on the effective transmission 
coefficient across the origin. 
We benchmark our analytic predictions against exact numerical simulations and
find excellent agreement in the hydrodynamic limit, apart from finite-time corrections. 
}

%############################################
\section{Introduction}
\label{sec:intro}

The dynamics of quantum many-body systems out of equilibrium is a central theme
in modern condensed matter physics and quantum information.
A paradigmatic scenario is a quantum quench~\cite{calabrese2016introduction}:
a system is prepared in an initial state that is not an eigenstate
of the Hamiltonian, and then allowed to evolve unitarily.
In one-dimensional integrable systems, the relaxation to a steady state
is constrained by an extensive set of local conservation laws,
and the ensuing dynamics is captured in terms of 
the underlying quasiparticles. Entanglement, transport properties  
and correlation functions are then expressed as sums
over these quasiparticles, leading to a hydrodynamic description
in the limit of large times and distances. 

When the Hamiltonian varies in space, the quasiparticles scatter off the
inhomogeneity, and the problem becomes that of a scattering problem in a
non-uniform medium. For smoothly varying backgrounds, the dynamics can be
described by Generalized Hydrodynamics (GHD)~\cite{bertini-2016,olalla-2016,bastianello2019generalized,alba2021generalized}. 
More abrupt inhomogeneities, such as defects and impurities, have been
extensively studied in the context of quantum transport. The seminal
Kane-Fisher problem~\cite{kane1992transport,kane1992transmission} set the stage for understanding
transport through a single impurity. Recent works addressed the
determination of the nonequilibrium steady state~\cite{bertini2016determination}, 
noninteracting transport in the presence of a defect~\cite{ljubotina-2019}, 
driven quantum point contacts~\cite{gamayun2021non,berdanier2019universal},
interacting defects~\cite{delvecchio2022transport}, and energy transport between
two integrable chains~\cite{biella2016energy}. The nonequilibrium steady state
generated by a moving defect was also studied~\cite{bastianello2018non}.
Systems of free fermions and bosons with a localized source were
analyzed in~\cite{krapivsky-2019} and~\cite{krapivsky-2020}, respectively, 
also in dissipative settings~\cite{alba2022noninteracting,tarantelli2022out}.

Entanglement dynamics in the presence of defects has attracted considerable
attention as well. The logarithmic growth of entanglement across a conformal
defect was first observed in critical chains~\cite{eisler2012on}. Exact
results for quenches in the presence of a conformal defect were also obtained
recently~\cite{capizzi2023entanglement}. Entanglement growth in wire junctions
was investigated in~\cite{calabrese2012entanglement}. The so-called modular Hamiltonian
for a free Dirac field with defects was derived in~\cite{mintchev2021modular}.
The time evolution of the entanglement negativity across a defect was
studied in~\cite{gruber2020time}, while the von Neumann entropy and the negativity
in fermionic chains with dissipative defects were solved exactly in~\cite{alba2021unbounded}
and~\cite{caceffo2023entanglement} (see also~\cite{fraenkel2021entanglement} for unitary
dynamics). 

Domain-wall initial states and inhomogeneous initial conditions provide another
class of problems where exact results are available. Domain wall melting across
a defect was studied in~\cite{capizzi2023domain}. The full counting statistics
for domain wall profiles was derived in~\cite{gamayun-2020}. 

\begin{figure}[t]
\centering
\includegraphics[width=.6\linewidth]{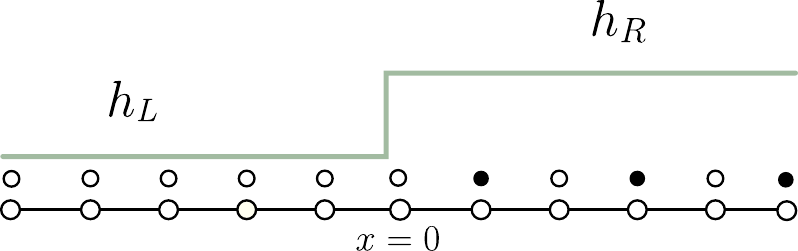}
\caption{ Setup employed in this work. We consider the inhomogeneous tight-binding chain 
	with two different magnetic fields (chemical potentials) $h_L$ and $h_R$ in the 
	two semi-infinite halves of the chain. We consider unitary dynamics from several 
	initial states. The figure shows the inhomogeneous N\'eel state in which the right-half 
	of the chain is prepared in the N\'eel state and the left part is in the fermionic 
	vacuum state. 
}
\label{fig:cartoon}
\end{figure}

Here we consider the paradigmatic setting of the XX chain with an inhomogeneous magnetic field. 
The setup is illustrated in Fig.~\ref{fig:cartoon}. The Hamiltonian of the system is 

\begin{equation}
\label{eq:xx-ham}
H = 
-\frac{J}{2} \sum_{j} \left( \sigma^+_j\sigma^-_{j+1}
 + \sigma^+_{j+1} \sigma^-_j \right) + h_L 
\sum_{j\leq 0} \sigma_j^z + h_R \sum_{j>0} \sigma_j^z, 
\end{equation}
where $\sigma_j^\pm$ are raising and lowering spin operators and $\sigma_j^{x,y,z}$ the 
standard Pauli matrices. In~\eqref{eq:xx-ham}, $J$ is a coupling constant, and $h_L,h_R$ are 
the magnetic fields in the two semi-infinite parts of the chain. 
Despite its interest as a paradigmatic setup to study dynamics under inhomogeneous Hamiltonians, very recently similar models   
attracted attention in the context of entanglement spreading and black-hole physics~\cite{saha2024generalised,ray2025page,kehrein2024page,li2025sharp}. 
Moreover, it should be possible to realize it experimentally with cold atoms in optical lattices~\cite{jepsen2020spin}. 
A similar setup was employed in Ref.~\cite{bucciantini2016probing} to study Klein tunneling in one-dimensional systems. 
Several properties of~\eqref{eq:xx-ham} can be derived in the hydrodynamic limit of large distances from the origin and long times 
by solving a stationary scattering problem~\cite{dipasquale2026entanglement}, 
yielding the effective reflection and transmission coefficients across the origin. For instance, this has recently allowed 
(see Ref.~\cite{dipasquale2026entanglement}) to conjecture a 
quasiparticle picture~\cite{calabrese2005evolution,fagotti2008evolution,alba2017entanglement,alba2018entanglement,klobas2021exact,bertini2022growth} 
 for entanglement spreading. The same strategy was applied to obtain the quasiparticle picture in the Kitaev chain 
with inhomogeneous fields~\cite{dipasquale2026entanglement}. 
Interestingly, while the generalized quasiparticle picture provided in Ref.~\cite{dipasquale2026entanglement} 
applies to quenches in the Kitaev chain starting from arbitrary 
initial states, in the XX chain it works only if one of the halves is prepared in the vacuum state. 
Clarifying this issue requires an \textit{ab initio} derivation of the quasiparticle picture, which in turn 
requires deriving the dynamics of the two-point correlation function. Here we obtain exact results 
for the time-dependent correlation function. Thus, besides its importance as a paradigmatic setup for dynamics 
under inhomogeneous Hamiltonians, the study of~\eqref{eq:xx-ham} is crucial to derive the complete quasiparticle 
picture for entanglement spreading. 

We exploit the fact that after a Jordan-Wigner transformation, Eq.~\eqref{eq:xx-ham} is mapped onto a 
fermionic quadratic Hamiltonian. Thus, the time-dependent two-point fermionic correlation function $G_{xy}=\langle c^\dagger_x c_y\rangle$ satisfies a  
linear system of equations. Although this system can be solved numerically, 
its analytic solution is challenging. Following Ref.~\cite{krapivsky-2019,viti-2016,alba2022noninteracting}, 
we provide a complete solution for the dynamics 
of the two-point correlation function after a quench in the inhomogeneous infinite $XX$ chain. 
We consider initial states that are product states, such as the domain-wall state, and the inhomogeneous N\'eel state (see Fig.~\ref{fig:cartoon}), 
in which the right half of the chain is prepared in the N\'eel state, and the left one in the ferromagnet with all 
the spins up (equivalently, the fermionic vacuum state). We also consider the quench from the homogeneous N\'eel state. 
The generalization to arbitrary initial states is straightforward. 
Following Ref.~\cite{krapivsky-2019}, we employ a convenient factorisation for the two-point correlator in terms of 
mode functions $S_{k,x}$, which allows us to 
map the problem to a single-particle scattering one. Then, by using a combination of Laplace and Fourier transforms,  
we convert this into a Riemann-Hilbert problem on the unit circle,
which we solve exactly. The correlator $G_{xy}$ is obtained by performing the inverse transforms, and it is written 
as an infinite sum 
over the mode functions.  Although the result is not expressed in terms of elementary functions, 
it allows us to rigorously derive the hydrodynamic limit of the correlation functions, 
valid when $x,y,t\to\infty$ with fixed ratios $x/t,y/t$.
In this limit, the correlation functions admit a  simple integral representation 
involving the reflection and transmission coefficients. 
We benchmark our analytic predictions against exact numerical simulations, 
finding excellent agreement in the hydrodynamic limit. 

%############################################
%############################################
%

The paper is organised as follows.
In Section~\ref{sec:model} we introduce the model and the scattering formalism.
In Section~\ref{sec:f-two} we present the linear system of equations describing the 
dynamics of the fermionic correlator. The solution of the system is discussed in 
Section~\ref{sec:RH}, Section~\ref{sec:inv-F}, and Section~\ref{sec:inv-laplace}. 
In Section~\ref{sec:hydro} we derive the hydrodynamic limit
and provide explicit formulas for the correlation functions.
In Section~\ref{sec:numerics} we compare our results with numerical data.
We conclude in Section~\ref{sec:conc} with a summary and outlook.

%############################################
\section{Mapping to the inhomogeneous tight-binding chain and stationary properties}
\label{sec:model}

Here we derive the dynamics of the fermionic two-point correlation function in the inhomogeneous 
XX chain. After a standard Jordan-Wigner transformation the XX Hamiltonian in~\eqref{eq:xx-ham} 
takes the form
\begin{equation}
	\label{eq:xx-ferm}
	H = H_L + H_R = -\frac{J}{2} \sum_{j} \left( c^\dagger_j c_{j+1} + c^\dagger_{j+1} c_j \right) + 
h_L \sum_{j\leq 0} c^\dagger_j c_j + h_R \sum_{j>0} c^\dagger_j c_j = \frac{1}{2} \sum_{j,l} c^\dagger_j H_{j,l} c_l,
\end{equation}
where $c_j$, $c^\dagger_j$ are standard Dirac fermions. In~\eqref{eq:xx-ferm} 
we defined the single-particle inhomogeneous Hamiltonian $H_{j,l}$ as
\begin{equation}
	\label{eq:xx-s-ferm}
	H_{j,l} = -\frac{J}{2} (\delta_{j+1,l} + \delta_{j-1,l}) +
\begin{cases}
h_L \delta_{j,l}, & j \le 0,\\
h_R \delta_{j,l}, & j > 0.
\end{cases}
\end{equation}

The left and right Hamiltonians $H_{L/R}$ in the thermodynamic limit can be diagonalised by going to Fourier space, obtaining
\begin{equation}
H_{L/R} = \int^{+\pi}_{-\pi} \frac{dk}{2\pi} \, \varepsilon_{L/R}(k) \, c^\dagger_k c_k,
\end{equation}
where $c_k = 1/L \sum_j e^{-ikj} c_j$ are Fourier transformed fermion operators. 
Notice that the Fourier modes $c_k$ are the same in the left and right parts of the 
chain. 
This is a peculiarity of the XX chain and would not be true for, e.g., the Ising model.
The dispersion relations $\varepsilon_{L/R}(k)$ read as
\begin{equation}
\varepsilon_{L/R}(k) = h_{L/R} - J \cos(k).
\end{equation}
The group velocities $v_{L/R}(k)$ are obtained as
\begin{equation}
	\label{eq:g-vel}
v_{L/R}(k) = \frac{d\varepsilon_{L/R}(k)}{dk} = J \sin(k),
\end{equation}
and are the same in the two parts of the system. 

It is useful for the following to compute the transmission probability across the 
interface. This can be done by solving  a single-particle scattering problem. 
Precisely, we consider a flux of right moving fermions approaching the interface 
from the left~\cite{dipasquale2026entanglement}. As ansatz for the scattering 
eigenfunctions, we consider the superposition of an incoming plane wave from the left with momentum $k_L > 0$ and of a reflected and transmitted waves with unknown amplitudes $R$ and $T$. Explicitly, the ansatz for the single-particle lattice scattering wavefunction $\Psi_j$ is
\begin{equation}
	\label{eq:se-ansatz}
	\Psi_j =
\begin{cases}
e^{i k_L j} + R e^{-i k_L j}, & j \le 0,\\
T e^{i k_R j}, & j > 0.
\end{cases}
\end{equation}
To proceed, we solve the Schr\"odinger equation $H_{j,l} \Psi_l = E \Psi_j$ $\forall j$, where we sum over $l$ and we fix the energy $E$. Importantly, since the magnetic fields are not the same in the left and right regions, the momentum $k_R > 0$ of the transmitted wave will be different from the one of the incoming wave $k_L$. In the bulk for $j < 0$ or $j > 1$, the ansatz~\eqref{eq:se-ansatz} satisfies the Schr\"odinger equation with energy
\begin{equation}
	\label{eq:e-cons}
\mu_L - J \cos(k_L) = \mu_R - J \cos(k_R) = E.
\end{equation}
Now, energy conservation~\eqref{eq:e-cons} can be used to rewrite the momentum $k_R$ of the transmitted wave in terms of $k_L$. We distinguish two different cases. If $1 \leq \cos(k_R) = \cos(k_L) + (h_R - h_L)/J \leq 1$, Eq.~\eqref{eq:e-cons} admits a real solution for $k_R$
\begin{equation}
	\label{eq:kr-sol}
	k_R = \arccos\left( \cos(k_L) + \frac{h_R}{J} - \frac{h_L}{J} \right),
\end{equation}
implying that transmission through the interface can occur. On the other hand, if $\cos(k_L) + (h_R - h_L)/t < -1$ or $\cos(k_L) + (h_R - h_L)/J > 1$ the equation~\eqref{eq:e-cons} has no real solutions for $k_R$. Physically, this corresponds to the fact 
that there is no asymptotic wave on the right with the same energy as 
the incoming one. In this case, the momentum $k_L$ is imaginary, 
corresponding to a evanescent wave, and the incoming quasiparticle undergoes 
total reflection. For $|h_R - h_L| \ge 2$ no modes can be transmitted through the origin. 
This implies that the transmission coefficient vanishes for all momenta and the linear 
growth of the entanglement entropy is absent. In the following we restrict ourselves to the situation with $1 \le \cos(k_R) = \cos(k_L) + (h_R - h_L)/J \le 1$. To compute the reflection $\mathcal{R}$ and transmission amplitudes $\mathcal{T}$, we need to impose the Schr\"odinger equation 
at the interface at $j = 0, 1$. After some algebra this yields the boundary conditions
\begin{equation}
	\label{eq:unit}
	1 + \mathcal{R} = \mathcal{T},
\end{equation}
\begin{equation}
	\label{eq:R-eq}
	e^{i k_L} + \mathcal{R} e^{-i k_L} = \mathcal{T} e^{i k_R},
\end{equation}
where $k_R$ is given in Eq.~\eqref{eq:kr-sol}. The solution of Eq.~\eqref{eq:unit} and~\eqref{eq:R-eq} are then
\begin{equation}
	\mathcal{R} = \frac{e^{i k_L} - e^{i k_R}}{e^{i k_R} - e^{-i k_L}} = \frac{\sin(k_L) - \sin(k_R) + i (h_R - h_L)/J}{\sin(k_L) + \sin(k_R) - i (h_R - h_L)/J},
\end{equation}
\begin{equation}
	\mathcal{T} = \frac{e^{i k_L} - e^{-i k_L}}{e^{i k_R} - e^{-i k_L}} = \frac{2\sin(k_L)}{\sin(k_L) + \sin(k_R) - i (h_R - h_L)/J},
\end{equation}
where we used Eq.~\eqref{eq:e-cons} to rewrite the expression in terms of $\sin(k_L)$ and $\sin(k_R)$. Finally we need to compute the reflection $R$ and transmission $T$ probabilities in terms of the amplitudes $\mathcal{R}$ and $\mathcal{T}$. The reflection probability is simply given by the absolute value squared of the amplitude $\mathcal{R}$
\begin{equation}
	R = |\mathcal{R}|^2 = \frac{(\sin(k_L) - \sin(k_R))^2 + (h_R - h_L)^2/J^2}{(\sin(k_L) + \sin(k_R))^2 + (h_R - h_L)^2/J^2}.
\end{equation}
On the other hand, since the velocity of the transmitted wave is different from the one of the incoming wave, the continuity equation for the probability current gives the transmission probability as
\begin{equation}
	\label{eq:T}
	T = \frac{v_R(k_R)}{v_L(k_L)} |\mathcal{T}|^2 = \frac{4\sin(k_L)\sin(k_R)}{(\sin(k_L) + \sin(k_R))^2 + (h_R - h_L)^2/J^2}.
\end{equation}

%############################################
\section{Dynamics of the fermionic two-point function}
\label{sec:f-two}

The outlined approach allows us to obtain the reflection and transmission amplitudes in the stationary situation, 
although the intermediate-time dynamics is not accessible. In the following we derive exact results for the dynamics of the 
two-point fermionic correlation matrix $G_{xy}$, which is defined as 
\begin{equation}
	\label{eq:Gxy}
	G_{xy}=\mathrm{Tr}(\rho(t)c^\dagger_xc_y), \quad \rho(t)=|\psi(t)\rangle\langle\psi(t)|. 
\end{equation}
By employing the Heisenberg equations of motion, since the Hamiltonian is quadratic, 
one obtains that $G_{xy}$ satisfies the linear system of equations 
\begin{multline}
\label{eq:one}
\frac{d G_{xy}}{dt}=\frac{i}{2}(G_{x+1,y}+G_{x-1,y}-G_{x,y+1}-G_{x,y-1})\\
+ih_L\Theta(-x-1)G_{xy}-ih_L\Theta(-y-1)G_{xy}+ih_R\Theta(x)G_{xy}-ih_R\Theta(y)G_{xy}, 
\end{multline}
where $\Theta(x)$ is the discrete Heaviside theta function and we defined $\Theta(0)=1$. 

Here we consider dynamics from initial states that are product states. Typical examples include the 
N\'eel state, which corresponds to 
\begin{equation}
	\label{eq:neel}
	G^{\mathrm{Neel}}_{xy}(0)=\delta_{x,y}, \quad x\,\,\textrm{even}.
\end{equation}
We also consider inhomogeneous initial states, such as the domain-wall initial state, which corresponds to 
\begin{equation}
	\label{eq:neel}
	G^{\mathrm{DW}}_{xy}(0)=\Theta(x)\delta_{x,y}.
\end{equation}
As outlined in Ref.~\cite{alba2022noninteracting}, having the solution of~\eqref{eq:one} for the N\'eel state 
allows one to obtain the solution for the inhomogeneous initial N\'eel defined by 
\begin{equation}
	G^{\mathrm{DNeel}}_{xy}(0)=\delta_{x,y}\Theta(x),\quad x\,\,\textrm{even}. 
\end{equation}
We should mention that in principle it is straightforward to obtain the dynamics starting 
from generic initial states, such as entangled ones. For instance, it should be possible to generalize our 
results to the dynamics starting from Fermi seas initial states, as discussed in Ref.~\cite{alba2022noninteracting}. 

To proceed we follow Ref.~\cite{alba2022noninteracting} (see also~\cite{krapivsky-2019} and~\cite{viti-2016}) 
and impose that the solution of~\eqref{eq:one} is factorized as 
\begin{equation}
\label{eq:ansatz}
G_{xy}=\sum_{k\in\mathbb{Z}}S_{k,x}\bar S_{k,y}, 
\end{equation}
where the bar denotes complex conjugation. 
Eq.~\eqref{eq:ansatz} is consistent with~\eqref{eq:one} provided that 
$S_{k,x}$ satisfies 
\begin{equation}
	\label{eq:S-evol}
	\frac{d S_{k,x}}{dt}=\frac{i}{2}[S_{k,x+1}+S_{k,x-1}] +iS_{k,x}(h_L\Theta(-x-1)+h_R\Theta(x)). 
\end{equation}
The initial conditions for the initial states discussed above are 
\begin{align}
	\label{eq:neel-state}
	 S^{\mathrm{Neel}}_{k,x}(0)&=\delta_{x,k},\quad k\,\,\textrm{even},\\
	 \label{eq:d-state}
	 S^{\mathrm{DW}}_{k,x}(0)&=\delta_{x,k}\Theta(k),\\
	 \label{eq:dneel-state}
	 S^{\mathrm{DNeel}}_{k,x}(0)&=\delta_{x,k}\Theta(k),\quad k\,\,\textrm{even}.
\end{align}
It is convenient to first perform a Laplace transform with 
respect to time in~\eqref{eq:S-evol}, defining $\widehat S_{k,x}$ as 
\begin{equation}
	\widehat S_{k,x}(s)=\int_0^\infty dt e^{-st}S_{k,x}(t). 
\end{equation}
This allows us to rewrite~\eqref{eq:S-evol} as 
\begin{equation}
\label{eq:laplace}
s \widehat S_{k,x}-S_{k,x}(0)=
\frac{i}{2}[\widehat S_{k,x+1}+\widehat S_{k,x-1}]+i\widehat{S}_{k,x}(h_L\Theta(-x-1)+h_R\Theta(x)). 
\end{equation}
Now, let us focus on the N\'eel state. It is clear that for odd $k$, Eq.~\eqref{eq:laplace} has 
the solution $\widehat S_{k,x}=0$. Similarly, in the domain wall case~\eqref{eq:d-state} 
one has $\widehat S_{k,x}=0$ for $k<0$. 
Moreover, it is convenient to consider the case with $S_{k,x}(0)=\delta_{k,x}$. Now, we have   
\begin{equation}
	\label{eq:lap-fou}
	s\widehat S_{k,x}-\delta_{k,x}=
	\frac{i}{2}[\widehat S_{k,x+1}+\widehat S_{k,x-1}]+
	i\widehat{S}_{k,x}(h_L\Theta(-x-1)+h_R\Theta(x)). 
\end{equation}
To proceed, we define the Fourier transformed $\widehat S_{k,q}$ as 
\begin{equation}
\label{eq:ft}
\widehat S_{k,q}=\sum_{x\in\mathbb{Z}}\widehat S_{k,x}e^{-iqx}, 
\end{equation}
where $q\in[-\pi,\pi]$. We also employ the representation of the discrete step function as  
\begin{equation}
	\Theta(x)=\int_{-\pi}^\pi\frac{dk}{2\pi}\frac{e^{ikx}}{1-e^{-i(k-i0^+)}},\quad \Theta(-x-1)=-\int_{-\pi}^\pi\frac{dk}{2\pi}\frac{e^{ikx}}{1-e^{-i(k+i0^+)}}, 
\end{equation}
where $0^+$ denotes a small positive real number. 
Finally, Eq.~\eqref{eq:lap-fou} becomes 
\begin{equation}
	\label{eq:lap-ff-1}
	s\widehat S_{k,q}-
	e^{-i q k}=i\widehat S_{k,q}\cos(q)+i\int_{-\pi}^\pi \frac{dp}{2\pi} \widehat S_{k,p}
\left[\frac{h_R}{1-e^{-i(q-p-i0^+)}}-\frac{h_L}{1-e^{-i(q-p+i0^+)}}\right]. 
\end{equation}
Eq.~\eqref{eq:lap-ff-1} is an integral equation for $\widehat S_{k,q}$. In the following, we show that 
the solution of~\eqref{eq:lap-ff-1} can be found by solving a ``simple'' Riemann-Hilbert problem. Notice also 
that a much simpler equation arises in the tight-binding chain with an impurity localized at the center of the chain~\cite{alba2022noninteracting}, 
which is recovered by removing the term in the square bracket in the right-hand side in~\eqref{eq:lap-ff-1}. In that case 
the solution of the equation is straightforward by integrating both members with respect to $q$. 
To introduce the Riemann-Hilbert problem, it is useful to introduce the projectors $(P_+ f)(z)$ 
\begin{align}
	\label{eq:proj-1}
	 (P_+f)(z)& =\oint\frac{dw}{2\pi i}\frac{f(w)}{w-z}, & |z|<1,\\
	\label{eq:proj-2}
	(P_-f)(z)& =-\oint\frac{dw}{2\pi i}\frac{f(w)}{w-z}, & |z|>1.
\end{align}
We can now rewrite the equations~\eqref{eq:lap-ff-1} as 
\begin{equation}
	\label{eq:ff-ll-2}
	s \widehat S_{k}-z^{k}=\frac{i}{2}\widehat S_{k}(z+1/z)+i[h_R(P_+ \widehat S_{k})+h_L(P_- \widehat S_{k})], 
\end{equation}
where  we defined $z=e^{-iq}$. In~\eqref{eq:ff-ll-2} we  removed the subscript $q$ in $\widehat S_{k,q}$ although it is clear 
that it depends on $q$  via $z$. Now, after using the Plemelj formula we can write the value of 
$\widehat S_{k}$ on the unit circle in terms of $(P_\pm \widehat S_{k})$ as 
\begin{equation}
	\widehat S_{k}= (P_+\widehat S_{k})+(P_-\widehat S_{k}). 
\end{equation}
After substituting in~\eqref{eq:ff-ll-2}, we obtain 
\begin{equation}
	\Big(s-\frac{i}{2}z-\frac{i}{2z}-ih_R\Big)\widehat S^+_{k}(z)+
	\Big(s-\frac{i}{2}z-\frac{i}{2z}-i h_L\Big)\widehat S^-_{k}-z^{k}
	=0,
\end{equation}
where we introduced the notation $\widehat S^\pm_{k}=(P_\pm \widehat S_{k})$, which finally yields
\begin{equation}
	\label{eq:RH-p}
	\widehat S^+_{k}(z)=-\frac{z^2+2(h_L+is)z+1}{z^2+2(h_R+is)z+1}\widehat S_{k}^-+\frac{z^k}{
		s-iz/2-1/(2z)-ih_R	
	}. 
\end{equation}
For convenience, we rewrite 
\begin{equation}
	\label{eq:fz}
	-\frac{z^2+2(h_L+is)z+1}{z^2+2(h_R+is)z+1}=-\frac{(z-l_-)(z-l_+)}{(z-r_-)(z-r_+)}:=f(z),
\end{equation}
where we introduced
\begin{equation}
	l_\pm= -h_L-is \pm\sqrt{(h_L+is)^2-1},
\end{equation}
and similarly $r_\pm$.

%############################################
\subsection{Solution of the associated Riemann-Hilbert problem}
\label{sec:RH}

%
%############################################
\begin{figure}[t]
\centering
\includegraphics[width=.4\linewidth]{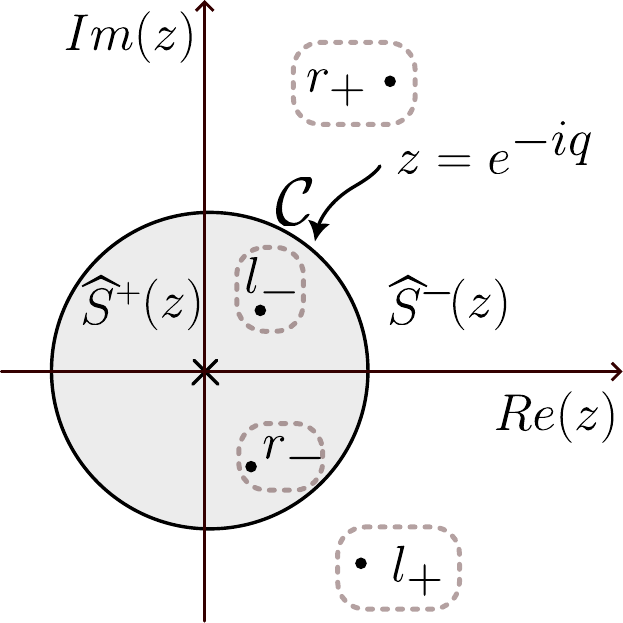}
\caption{ Riemann-Hilbert problem $\widehat S^+(z)=f(z)\widehat S^-(z)+g(z)$ (cf.~\eqref{eq:RH-p}) 
	for the functions $\widehat S^\pm(z)=\widehat S^\pm_{k,q}$, on  the unit circle $\mathcal{C}$ 
	with $z=e^{-iq}$. The functions $f(z)$ and $g(z)$ are defined in~\eqref{eq:RH-p}. 
	We denote with $l_\pm$ and $r_\pm$ the zeros and the poles of $f(z)$, 
	respectively. 
}
\label{fig:RH}
\end{figure} 
%############################################
%

Now, Eq.~\eqref{eq:RH-p} defines a so-called Riemann-Hilbert problem~\cite{gakhov1990boundary} for 
a function $\widehat S_{k}$. The boundary of the RH problem is the unit circle $\mathcal{C}$ (see Fig.~\ref{fig:RH}), 
on which~\eqref{eq:RH-p} has to be satisfied. To solve the RH problem, one first focuses 
on the homogeneous one, i.e., the one in which the last term in~\eqref{eq:RH-p} is set to 
zero. The solution $X(z)$ of the homogeneous problem can be written as~\cite{gakhov1990boundary}  
\begin{equation}
	\label{eq:RH-sol}
	X(z)=\exp\left(\oint \frac{d\zeta}{2\pi i}\frac{\ln f(\zeta)}{\zeta-z}\right), 
\end{equation}
where $f(z)$ is the function defined in~\eqref{eq:fz}. 
Notice that in Eq.~\eqref{eq:RH-sol} we assume a specific choice for the branch cut of the logarithm function, for 
which $\ln(f(z))$ is analytic on the unit circle. This amounts to choosing a branch cut that lies within the unit 
circle. Since the poles and zeros of $f(z)$ come in pairs this is always possible. 
Precisely, since $l_+l_-=1$ and $r_+r_-=1$, the number of zeros of $f(z)$ inside the unit circle 
$\mathcal{C}$ equals the number of poles this choice is always possible. 
The solution depends on the position of the zeroes and the poles of the function $f(z)$ (cf.~\eqref{eq:fz}). Precisely, 
let us discuss the behavior of $l_\pm$ and $r_\pm$ as a function of complex $s$. 
Let us restrict ourselves to $h_L,h_R>0$. 
One can straightforwardly verify that $r_+$ and $l_+$ are inside the unit circle $\mathcal{C}$ for $\mathrm{Im}(s)<h_R$ and 
$\mathrm{Im}(s)<h_L$, respectively. Moreover, one has that $l_-$ and $r_-$ are inside $\mathcal{C}$ for 
$\mathrm{Im}(s)>h_R$ and $\mathrm{Im}(s)>h_L$, respectively.  
Now, let us consider, without loss of generality, the situation with $h_L>h_R$. This means that there are three 
cases to consider depending on the value of $\mathrm{Im}(s)$. Precisely, we have to treat separately the case with $\mathrm{Im}(s)>h_L$, 
with $h_R<\mathrm{Im}(s)\le h_L$, and $\mathrm{Im}(s)<h_R$. 
To illustrate the solution of the Riemann-Hilbert problem, let us consider the case with 
$\mathrm{Im}(s)>h_L$. Now, one has that  
$l_-,r_-$ are within the unit circle. By using~\eqref{eq:RH-sol}, the solution of the homogeneous problem $X(z)$ is 
\begin{align}
	X(z)&=\frac{z-l_+}{z-r_+}\Theta(1-|z|)-\frac{z-r_-}{z-l_-}\Theta(|z|-1), & |l_-|<1,\,|r_-|<1.
\end{align}
From the solution $X(z)$ of the homogeneous problem, one obtains the solution $\widehat S^{\pm}(z)$ of 
the inhomogeneous one as 
\begin{equation}
	\label{eq:RH-inh-sol}
	\widehat S_k^\pm(z)=X^\pm(z)\left(\oint\frac{d\zeta}{2\pi i}\frac{g(\zeta)}{X^+(\zeta)(\zeta-z)}\right), 
\end{equation}
where $X(z)$ is obtained from~\eqref{eq:RH-sol}. 
Notice that in the integral, one has $X^+(\zeta)$ for both $\widehat S_k^+(z)$ and 
$\widehat S_k^-(z)$. Here $g(z)$ is given as 
\begin{equation}
	g(z)=\frac{z^k}{s-iz/2-i/(2z)-ih_R}. 
\end{equation}
The solution on the unit circle is 
\begin{equation}
	\widehat S_k(z)=\widehat S_k^+(z)+\widehat S_k^-(z),\quad z\in\mathcal{C}. 
\end{equation}
A straightforward application of Cauchy' residue theorem to~\eqref{eq:RH-inh-sol} 
gives 
%
%\boxed{
\begin{align}
\label{eq:Sp-ds}
& \widehat S^+_{k,j}(z) = \frac{2i z^{k+1}}{(z-r_-)(z-r_+)}
- \frac{2i r_{a_j}^{k+1}(z-l_{b_j})}{(z-r_-)(z-r_+)(r_{a_j}-l_{b_j})},
\qquad |z|<1,\; k\ge -1,\\
\label{eq:Sm-ds}
& \widehat S^-_{k,j}(z) = \frac{2i r_{a_j}^{k+1}}{(z-l_{c_j})(r_{a_j}-l_{b_j})},
\qquad |z|>1,\; k\ge -1,
\end{align}
where the indices $(a_j,b_j,c_j)$ and the corresponding region of $\operatorname{Im}(s)$ are given in Table~\ref{tab:tab0}.
\begin{table}[t]
\[
\begin{array}{c|c|c|c|c}
j & \text{Region for } \operatorname{Im}(s) & a_j & b_j & c_j \\ \hline
1 & \mathrm{Im}(s)> h_L & - & + & - \\
2 & h_R < \mathrm{Im}(s) \le h_L & - & - & + \\
3 & \mathrm{Im}(s)\le h_R & + & - & +
\end{array}
\]
\caption{Definitions of $a_j,b_j,c_j$ appearing in~\eqref{eq:Sp-ds} and~\eqref{eq:Sm-ds}. The domain 
	$\mathrm{Im}(s)$ for different values of $j$ is also reported. 
}
\label{tab:tab0}
\end{table}
%
%
%\begin{align}
%	\label{eq:Sq-1}
%	& \widehat S_{k,1}^+(z)=\frac{2i z^{k+1}}{(z-r_-)(z-r_+)}-
%	\frac{2i r_-^{k+1}(z-l_+)}{(z-r_-)(z-r_+)(r_--l_+)} & |z|<1,\,k\ge -1\\
%	\label{eq:Sq-2}
%	& \widehat S_{k,1}^-(z)=
%	\frac{2i r_-^{k+1}}{(z-l_-)(r_--l_+)} & |z|>1,\,k\ge -1
%\end{align}
%
%In the region $h_R\le \mathrm{Im}(s)\le h_L$, a similar calculation gives 
%
%\begin{align}
%	\label{eq:Sq-1x}
%	& \widehat S_{k,2}^+(z)=\frac{2i z^{k+1}}{(z-r_-)(z-r_+)}-
%	\frac{2i r_-^{k+1}(z-l_-)}{(z-r_-)(z-r_+)(r_--l_-)} & |z|<1,\,k\ge -1\\
%	\label{eq:Sq-2x}
%	& \widehat S_{k,2}^-(z)=
%	\frac{2i r_-^{k+1}}{(z-l_+)(r_--l_-)} & |z|>1,\,k\ge -1
%\end{align}
%
%Finally, for $\mathrm{Im}(s)<h_R$, one obtains 
%
%\begin{align}
%	\label{eq:Sq-1xx}
%& \widehat S_{k,3}^+(z)=\frac{2i z^{k+1}}{(z-r_-)(z-r_+)}-
%	\frac{2i r_+^{k+1}(z-l_-)}{(z-r_-)(z-r_+)(r_+-l_-)} & |z|<1,\,k\ge -1\\
%	\label{eq:Sq-2xx}
%	& \widehat S_{k,3}^-(z)=
%	\frac{2i r_+^{k+1}}{(z-l_+)(r_+-l_-)} & |z|>1,\,k\ge -1
%\end{align}
%
To obtain $\widehat S_{k}(z)$ for $k<-1$, it is convenient to change variable as 
$\zeta\to1/\zeta$ in~\eqref{eq:ff-ll-2}. We do not show the derivation of the results, 
reporting the final result for $S_{k,x}$ in Section~\ref{sec:inv-laplace}. 

%############################################
\subsection{Performing the inverse Fourier transform}
\label{sec:inv-F}

To proceed, let us perform the inverse Fourier transform of $\widehat S^\pm_{j}$ 
with respect to $q$ (equivalently $z=e^{-iq}$). 
The inverse Fourier transform can be performed analytically after observing that 
\begin{equation}
	\label{eq:lap-1}
	\int_{-\pi}^\pi\frac{dq}{2\pi}\frac{e^{-i q (x-k)}}{h_R+is+\cos(q)}
	=-\frac{r_-^{|x-k|}}{\sqrt{(h_R+is)^2-1}}. 
\end{equation}
Eq.~\eqref{eq:lap-1} allows one to straightforwardly obtain the inverse Fourier transform of the first term in~\eqref{eq:Sp-ds}. 
Moreover, Eq.~\eqref{eq:lap-1} allows us to derive the inverse Fourier transform of all the remaining terms. 
For instance, the second term in~\eqref{eq:Sp-ds} for $j=1$ is 
\begin{equation}
	\label{eq:second}
	-\frac{2i r_-^{k+1}(z-l_+)}{(z-r_-)(z-r_+)(r_--l_+)}=-\frac{i r_-^{k+1}(e^{iq}-l_+)}{r_--l_+}\frac{e^{-iq}}{h_R+is+\cos(q)}.
\end{equation}
We can rewrite the denominator in the right-hand side of~\eqref{eq:second} 
in two different ways as  
\begin{align}
	& \frac{1}{r_--l_+}=-\frac{1}{2(h_L-h_R)}\frac{l_+-r_+}{l_+},\\
	& \frac{1}{r_--l_+}=\frac{1}{2(h_R-h_L)}\frac{l_--r_-}{r_-}.
\end{align}
This allows us to rewrite~\eqref{eq:second} as 
\begin{equation}
	-\frac{2i r_-^{k+1}(z-l_+)}{(z-r_-)(z-r_+)(r_--l_+)}=\frac{i r_-^k
	\left[(l_--r_-)+
	 r_-(l_+-r_+)e^{-iq}\right] 
	}{2(h_R-h_L)(h_R+is+\cos(q))}.
\end{equation}

\begin{table}[t]
\centering
\caption{Inverse Fourier transform of~\eqref{eq:Sp-ds} and~\eqref{eq:Sm-ds}. The results for 
$\widehat S^+_{k,j}$ have to be multiplied by $1/(r_+-r_-)$. The index $j=1,2,3$ labels the three 
cases for $\operatorname{Im}(s)$ as defined in the text.}
\label{tab:invFTcombined}
\small
\setlength{\tabcolsep}{2pt}
\begin{tabular}{c >{\raggedright\arraybackslash}p{7.2cm} >{\raggedright\arraybackslash}p{6.2cm}}
\toprule
$j$ & $ (r_+-r_-)\widehat S^+_{k,j}(x)$ & $\widehat S^-_{k,j}(x)$ \\
\midrule
1 &
$
\begin{array}{@{}l@{}}
-2i r_-^{|x-k|} \\[2pt]
+ \dfrac{i r_-^k}{h_R-h_L}
\Bigl[ r_-^{|x|}(l_- - r_-) + r_-(l_+ - r_+)r_-^{|x+1|} \Bigr]
\end{array}$ &
$\dfrac{i}{h_R-h_L} r_-^k l_-^{-x-1}(l_- - r_-)\,\Theta(-x)$ \\
\addlinespace
2 &
$
\begin{array}{@{}l@{}}
-2i r_-^{|x-k|} \\[2pt]
+ \dfrac{i r_-^k}{h_R-h_L}
\Bigl[ r_-^{|x|}(l_+ - r_-) - l_-(l_+ - r_-)r_-^{|x+1|} \Bigr]
\end{array}$ &
$\dfrac{i}{h_R-h_L} r_-^k l_+^{-x-1}(l_+ - r_-)\,\Theta(-x)$ \\
\addlinespace
3 &
$
\begin{array}{@{}l@{}}
-2i r_+^{|x-k|} \\[2pt]
- \dfrac{i r_+^{k+1}}{h_R-h_L}
\Bigl[ r_+^{|x|}(r_- - l_-)l_+ - (r_- - l_-) r_+^{|x+1|} \Bigr]
\end{array}$ &
$\dfrac{i}{h_R-h_L} r_+^{k+1} l_+^{-x}(r_- - l_-)\,\Theta(-x)$ \\
\bottomrule
\end{tabular}
\end{table}
To perform the inverse Laplace transform one has to consider the branch cuts introduced by 
the square root functions in $r_\pm$ and $l_\pm$. Now, $r_\pm$ have zeros at $s=i(h_R\pm 1)$, i.e., along 
the imaginary axis. Similarly, $l_\pm$ vanish at $s=i (h_L\pm 1)$. We choose the branch cuts of the 
square root in $[i\min(h_L+1,h_L-1),i\max(h_L+1,h_L-1)]$. The resulting branch cut for the function $\widehat S_{kq}(s)$ is 
$[i\min(h_L-1,h_R-1),i\max(h_L+1,h_R+1)]$, i.e., the ``superposition'' of the two cuts. 

%############################################
\subsection{Inverse Laplace transform} 
\label{sec:inv-laplace}

To obtain $S_{k,x}(t)$ we have to perform the inverse Laplace transform of $\widehat S^\pm_{k,j}(x)$ (see Table~\ref{tab:invFTcombined}), which is defined as 
\begin{equation}
	\label{eq:inv-lap}
	S_{k,x}(t)=\int_{\gamma-i\infty}^{\gamma+i\infty}\frac{ds}{2\pi i} e^{st} \widehat S_{k}(s), 
\end{equation}
where $\gamma$ is large enough to avoid singularities. 
After putting together the different contributions reported in Table~\ref{tab:invFTcombined}, 
the final result for $\widehat S_{k,x}(s)$ is for $k\ge -1$
\begin{multline}
\label{eq:pos-k}
		\widehat S_{k,x}(s)=\frac{2i}{r_+-r_-}\left\{r_+^{|x-k|}+\frac{r_+^{k+x}(r_+-l_+)+r_+^{k+x+1}(1-r_+l_-)}{2(h_R-h_L)}
		\right\}\Theta^+(x)\\
		+\frac{i}{h_R-h_L}r_+^kl_+^{-x-1}(l_+-r_+)\Theta^+(-x), 
\end{multline}
where $\Theta^+(x)$ is 
the standard Heaviside theta function with the shifted argument as $x\to x+0^+$. 
Moreover, we verified that for  $k<-1$ the result is given by 
\begin{equation}
	\label{eq:neg:k}
	\widehat S_{k,x}(s)=\left\{ \begin{array}{cc}
			\left. \widehat S_{-x-1,-k-1}(s)\right|_{h_R\leftrightarrow h_L}, & x\le-1,\\
			\left. \widehat S_{-k-1,-x-1}(s)\right|_{h_R\leftrightarrow h_L}, & x>-1.
	\end{array}\right.
\end{equation}
%
%
%############################################
\begin{figure}[t]
\centering
\includegraphics[width=.6\linewidth]{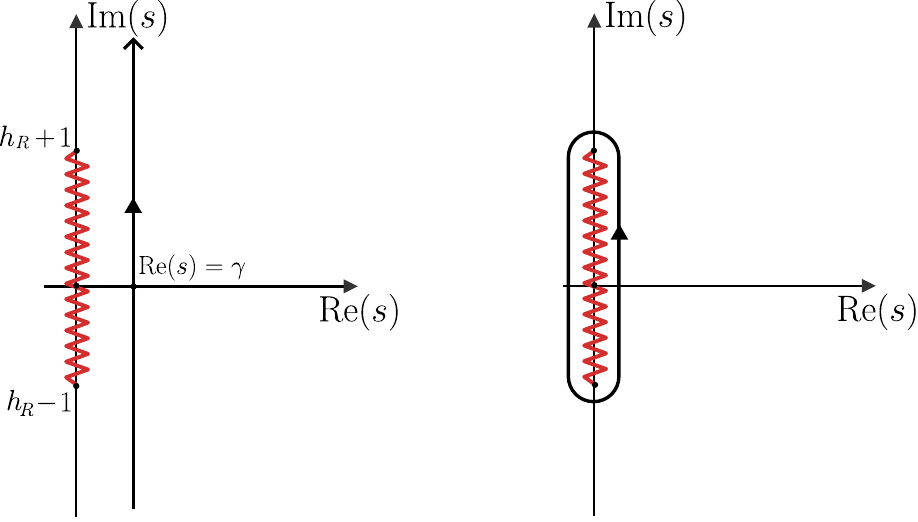}
\caption{ (Left) Contour employed to compute the inverse Laplace transform in~\eqref{eq:inv-lap}. The wiggly vertical 
	line is a branch cut, and $\gamma$ is a positive real part to avoid singularities. (Right) The contour in the 
	left panel is deformed to wrap around the branch cut. 
}
\label{fig:contour}
\end{figure} 
%############################################
%
Now, to perform the inverse Laplace transform in~\eqref{eq:inv-lap} we can deform the contour as illustrated in  
Fig.~\ref{fig:contour}. 
The integral gets contributions only from the regions in $s$ near the branch cuts of the square root functions. 
Precisely, the term $r_\pm$ have a branch cut in $i[h_R-1,h_R+1]$, whereas $l_\pm$ has a branch cut in $i[h_L-1,h_L+1]$. 
Thus, it is convenient to split the integration domain. We have to distinguish the situation with $h_L-h_R>2$ (recalling that we assume $h_L>h_R$) 
and $h_L-h_R<2$. In the former case there is no overlap between the branch cuts of $r_\pm$ and $l_\pm$. This corresponds 
to the situation in which there is no transport between the two parts of the chain, and we do not discuss it. 
Let us consider the case $h_L-h_R\le 2$. We can split the integration domain as $i[h_R-1,h_L-1]\cup i[h_L-1,h_R+1]\cup i[h_R+1,h_L+1]$. 

It is convenient to rewrite the inverse Laplace transform for $k\ge -1$ as sum of three terms  
\begin{equation}
	\label{eq:S-dec}
	S_{k,x}(t)=S^{(1)}_{k,x}(t)+S^{(2)}_{k,x}(t)+S^{(3)}_{k,x}(t).
\end{equation}
After defining the three intervals $\mathcal{C}_j$ as 
% Define the integration intervals
\[
\mathcal C_1 = [1-h_L,\; 1-h_R], \qquad
\mathcal C_2 = [-1-h_R,\; 1-h_L], \qquad
\mathcal C_3 = [-1-h_L,\; -1-h_R],
\]
the terms $S_{k,x}^{\scriptscriptstyle (j)}$ in \eqref{eq:S-dec} take the form 
% Unified compact formula for S^{(j)}
\begin{multline}
\label{eq:Sj-compact}
S^{(j)}_{k,x}(t) = 
\int_{\mathcal C_j} \frac{ds}{2\pi} e^{-ist}
\left[
\frac{2i}{\rho_+-\rho_-}
\left( \rho_+^x \, \mathcal A_+^{(j)} + \rho_-^x \, \mathcal A_-^{(j)} \right) \Theta^+(x)\right.\\
\left.\;+\;
\frac{i}{h_R-h_L}
\left( \rho_+^k \lambda_{\alpha_{j}}^{-x-1}(\lambda_{\alpha_{j}}-\rho_+)
- \rho_{\beta_{4-j}}^k \lambda_{\beta_{j}}^{-x-1}(\lambda_{\beta_{j}}-\rho_{\beta_{4-j}}) \right) \Theta^+(-x)
\right].  
\end{multline}
Here we introduced 
\begin{align}
	\label{eq:rho}
	& \rho_{\pm}=-h_R-s\pm\sqrt{(h_R+s)^2-1},\\
	\label{eq:lambda}
	&\lambda_\pm=-h_L-s\pm\sqrt{(h_L+s)^2-1},
\end{align}
and the coefficients $\mathcal{A}_\pm^{\scriptscriptstyle(j)}$ are defined in Table~\ref{tab:tab1}. 
The result for $k<-1$ is obtained from the results above and~\eqref{eq:neg:k}. One has to observe that the integration domain 
has to be changed as $\mathcal{C}_1\leftrightarrow \mathcal{C}_3$, whereas $\mathcal{C}_2$ remains the same. 

%############################################
\section{Hydrodynamic limit}
\label{sec:hydro}

\begin{table}[t]
\centering
\small
\setlength{\tabcolsep}{2pt}
\begin{tabular}{c >{\raggedright\arraybackslash}p{6.2cm} >{\raggedright\arraybackslash}p{6.2cm} c c}
\toprule
$j$ & $\mathcal A_+^{(j)}$ & $\mathcal A_-^{(j)}$ & $\alpha_j$ & $\beta_j$ \\
\midrule
1 &
$\begin{array}{@{}l@{}}
\rho_+^{-k} + \dfrac{\rho_+^k(\rho_+-\lambda_+)+\rho_+^{k+1}(1-\lambda_-\rho_+)}{2(h_R-h_L)}
\end{array}$ &
$\begin{array}{@{}l@{}}
\rho_-^{-k} + \dfrac{\rho_-^k(\rho_- - \lambda_+)+\rho_-^{k+1}(1-\lambda_-\rho_-)}{2(h_R-h_L)}
\end{array}$ &
$+$ & $+$ \\
\addlinespace
2 &
$\begin{array}{@{}l@{}}
\rho_+^{-k} + \dfrac{\rho_+^k(\rho_+-\lambda_+)+\rho_+^{k+1}(1-\lambda_-\rho_+)}{2(h_R-h_L)}
\end{array}$ &
$\begin{array}{@{}l@{}}
\rho_-^{-k} + \dfrac{\rho_-^k(\rho_- - \lambda_-)+\rho_-^{k+1}(1-\lambda_+\rho_-)}{2(h_R-h_L)}
\end{array}$ &
$+$ & $-$ \\
\addlinespace
3 &
$\begin{array}{@{}l@{}}
\dfrac{\rho_+^{k}(\lambda_- - \lambda_+)-\rho_+^{k+2}(\lambda_- - \lambda_+)}{2(h_R-h_L)}
\end{array}$ &
$0$ &
$+$ & $-$ \\
\bottomrule
\end{tabular}
\label{tab:tab1}
\caption{Coefficients $\mathcal A_\pm^{(j)}$ and indices $\alpha_{j}$ and $\beta_j$ for $j=1,2,3$ appearing 
in Eq.~\eqref{eq:Sj-compact}. The functions $\rho_\pm$ and $\lambda_\pm$ are defined in~\eqref{eq:rho} and~\eqref{eq:lambda}. 
}
\end{table}
The results of the previous section allow one to obtain analytically the dynamics of the 
fermionic two-point function by using~\eqref{eq:ansatz}, where one has to perform the sum over 
$k$ and the integrals in~\eqref{eq:Sj-compact} numerically. In the following 
we derive the dynamics of $G_{xy}(t)$ in the hydrodynamic limit $x,y,t\to\infty$ with fixed $x/t,y/t$. 
The result depends on whether $x,y$ are on the same side of the origin or in different ones, and on the 
sign of $k$ (cf.~\eqref{eq:ansatz}). 
Let us first isolate the terms with positive $k$ restricting also to the situation with $x<0,y<0$ 
in~\eqref{eq:ansatz}, defining $G^{\scriptscriptstyle [+--]}_{xy}$ as
\begin{equation}
G_{xy}^{\scriptscriptstyle [+--]}(t)=\sum_{k=0}^\infty S_{k,x}\bar S_{k,y}. 
\end{equation}
$G_{xy}^{\scriptscriptstyle[+--]}$ is the fermion correlator for $x,y<0$ for the quench in which the right part of the 
chain is prepared in the fully-occupied state (cf.~~\eqref{eq:d-state}). By restricting the sum to even $k$, one 
obtains the result for the quench from~\eqref{eq:dneel-state}. In the following we provide results for 
the quenches in which only the left part or the right part of the chain are prepared in the N\'eel state. 
The definition of the generic $G_{xy}^{\scriptscriptstyle [\sigma_1\sigma_2\sigma_3]}$ with $\sigma_j=\pm$ 
follows straightforwardly. One can verify that in the hydrodynamic limit one has  
\begin{multline}
	\label{eq:S2-int}
	G_{xy}^{\scriptscriptstyle[+--]}=
\int_{-1-h_R}^{1-h_L}\frac{ds}{2\pi}\int_{-1-h_R}^{1-h_L}\frac{ds'}{2\pi}
e^{-i(s-s')t-i(x+1) A_L(s)+i(y+1)A_L(s')-i\pi(x-y)}
\\\times	\sum_{k=0}^\infty  e^{ik(A_R(s)-A_R(s'))}
\frac{(\lambda_+(s)-\rho_+(s))(\bar\lambda_+(s')-\bar\lambda_+(s'))}{(h_R-h_L)^2}, 
\end{multline}
where we defined 
\begin{equation}
	A_{L/R}(s)=-\arccos(h_{L/R}+s), 
\end{equation}
and the functions $\lambda_{\pm}$ and $\rho_\pm$ are defined in~\eqref{eq:rho} and~\eqref{eq:lambda}. 
Eq.~\eqref{eq:S2-int} is obtained from~\eqref{eq:Sj-compact}, i.e., neglecting the contributions for $j=1,3$
because they are suppressed in the hydrodynamic limit. Moreover, by employing the stationary 
phase approximation one can show that in the product 
$S^{\scriptscriptstyle(2)}_{k,x}S_{k,y}^{\scriptscriptstyle(2)}$ 
the terms containing time-dependent or position-dependent phase factors in the 
limit $s\to s'$ can be neglected. After performing the sum over $k$ and changing variables as 
$\arccos(s+h_L)\to s$ and $\arccos(s'+h_L)\to s'$, we obtain 
\begin{multline}
	\label{eq:G-int}
	G^{\scriptscriptstyle[+--]}_{x,y}=
	\int_0^{K_\mathrm{max}}\frac{ds}{2\pi}\frac{ds'}{2\pi}
e^{-i(\cos(s)-\cos(s'))t+i(x+1)s-i(y+1)s'-i\pi(x-y)}
\\\frac{\sin(s)\sin(s')}{1-e^{i[A'_R(s)-A'_R(s')]}}
\frac{(\lambda'_+(s)-\rho'_+(s))(\bar\lambda'_+(s')-\bar\lambda'_+(s'))}{(h_R-h_L)^2}, 
\end{multline}
where we defined $K_\mathrm{max}=\arccos(-1-h_R+h_L)$. In~\eqref{eq:G-int} the prime in the integrand is 
to stress that the functions have to be modified to take into account the change of variables, e.g., 
$A_{L}'(s)=A_L(\cos(s)-h_L)$. Again, we should stress that we have the conditions $h_R<h_L$ and $h_L-h_R<1$ 
with both $h_L$ and $h_R$ positive. In the hydrodynamic limit the behavior of the correlator is 
dominated by the contribution of the pole at $s\to s'$. Thus, it is convenient to change variables as 
\begin{equation}
	Q=s-s',\quad K=\frac{s+s'}{2}. 
\end{equation}
After changing variables, we obtain 
\begin{multline}
	\label{eq:den-int}
	G_{xy}^{\scriptscriptstyle[+--]}=
\int_0^{K_{\mathrm{max}}}\frac{dK}{2\pi}\int_{-2\min(K,K_\mathrm{max}-K)}^{2\min(K,K_\mathrm{max}-K)}
\frac{dQ}{2\pi}
e^{-it(\cos(K+Q/2)-\cos(K-Q/2))}\\
\times\frac{(\lambda'_+(K+Q/2)-\rho'_+(K+Q/2))(\bar\lambda'_+(K-Q/2)-\bar\lambda'_+(K-Q/2))}{(h_R-h_L)^2}\\
\times e^{i(K+Q/2)(x+1)-i(K-Q/2)(y+1)-i\pi(x-y)}
\frac{\sin(K+Q/2)\sin(K-Q/2)}{1-e^{i[A'_R(K+Q/2)-A'_R(K-Q/2)]}}. 
\end{multline}
We can Taylor expand the integrand in the limit $Q\to0$. We can also perform the integral over $Q$ by using that 
\begin{equation}
	\int_{-\infty}^\infty\frac{dQ}{2\pi i} \frac{e^{i Q x}}{Q\pm i0^+}=\Theta(\mp x), 
\end{equation}
where the regulator $0^+$ is put to obtain a finite result. 
Finally, we obtain  
\begin{equation}
	\label{eq:Gxy-pmm}
	G^{\scriptscriptstyle[+--]}_{x,y}(t)=\int_0^{\pi}\frac{dk}{2\pi}
	\mathrm{Re}[T(k)]e^{i(k+\pi)(x-y)}\Theta((x+y+2)/2+tv(k))\Theta(-x)\Theta(-y), 
\end{equation}
where $T(k)$ is the transmission coefficient~\eqref{eq:T}. In~\eqref{eq:Gxy-pmm} 
$\mathrm{Re}[T(k)]$ is nonzero only for $k\in[0,K_\mathrm{max}]$. 
Notice the presence of the staggering phase $\pi(x-y)$. In~\eqref{eq:Gxy-pmm}   
we introduced the theta functions $\Theta(-x)\Theta(-y)$ to stress that the result holds for negative $x,y$. 
Finally, let us observe that for the quench in which the right part of the chain is prepared in the N\'eel state and 
the left one is empty (cf.~\eqref{eq:dneel-state}) one has to sum over even $k$ only in~\eqref{eq:ansatz}. 
The derivation remains the same, the only change is that one has $2(A'_R(s)-A'_R(s'))$ in the 
phase in the denominator in~\eqref{eq:G-int}. The result is half of~\eqref{eq:Gxy-evol}. 

Let us now consider the result for $x,y>0$, i.e., $G_{xy}^{\scriptscriptstyle[+++]}$. 
To proceed, we start observing that in the hydrodynamic limit 
$x,y,t\to\infty$ the term with $j=3$ in~\eqref{eq:Sj-compact} vanishes, and 
only the terms with $j=1,2$ contribute. They correspond to left and right moving fermions in the right part of the 
chain. 
Moreover, one can neglect the terms  of the form $S^{\scriptscriptstyle(\alpha)}_{k,x}\bar S_{k,y}^{\scriptscriptstyle(\beta)}$, 
with $\alpha\ne\beta$. By employing the stationary phase approximation one can verify that only terms of the form $\rho_+^x\bar\rho_+^y$ 
and $\rho_+^{-x}\bar \rho_+^{-y}$ contribute. The derivation is similar to that of $G_{xy}^{\scriptscriptstyle[+--]}$. The 
final result is 
\begin{multline}
	\label{eq:Gxy-ppp}
	G^{\scriptscriptstyle[+++]}_{xy}(t)=\frac{1}{2}\int_0^{\pi}\frac{dk}{2\pi} 
	\Big[e^{i\pi(x-y)-i k(x-y)}+e^{-i\pi(x-y)+ik(x-y)}\\-
e^{i\pi(x-y)-ik(x-y)}
\Theta(t v(k)-(x+y)/2)\mathrm{Re}[T(\pi-k)]\Big)
\Big]\Theta(x)\Theta(y),
\end{multline}
The first two terms in~\eqref{eq:Gxy-ppp} correspond to the right-moving and left-moving 
fermions. The last term accounts for the fermions that are not in the right-part of the chain 
because they are transmitted in the left part. 
Finally, the correlator for the case with $k<0$ in~\eqref{eq:ansatz} is obtained formally by applying~\eqref{eq:neg:k}. 
However, in the hydrodynamic limit the result is obtained by swapping the left and right parts of the chain. 
Precisely, for positive $x,y$ one has to change $x,y\to -x,-y$ and $h_L\leftrightarrow h_R$ in~\eqref{eq:Gxy-pmm}. 
One obtains 
\begin{equation}
	\label{eq:Gxy-evol}
	G^{\scriptscriptstyle[-++]}_{xy}(t)=\int_{0}^\pi\frac{dk}{2\pi}
	\mathrm{Re}[T(\pi-k)]e^{-i(k+\pi)(x-y)}\Theta(-(x+y+2)/2+tv(k))\Theta(x)\Theta(y), 
\end{equation}
where $T(\pi-k)$ is obtained from $T(k)$ when exchanging $h_L$ and $h_R$. 
Similarly, for negative $x,y$  we have 
\begin{multline}
	\label{eq:Gxy-mmm}
	G_{xy}^{\scriptscriptstyle[---]}(t)=\int_{0}^\pi\frac{dk}{2\pi}\Big[e^{ik(x-y)-i\pi(x-y)}
	+e^{-ik(x-y)+i\pi(x-y)}\\
-\Theta((x+y)/2+tv(k))e^{ik(x-y)-i\pi(x-y)}\mathrm{Re}[T(k)]\Big]\Theta(-x)\Theta(-y). 
\end{multline}
%
%############################################
\begin{figure}[t]
\centering
\includegraphics[width=.85\linewidth]{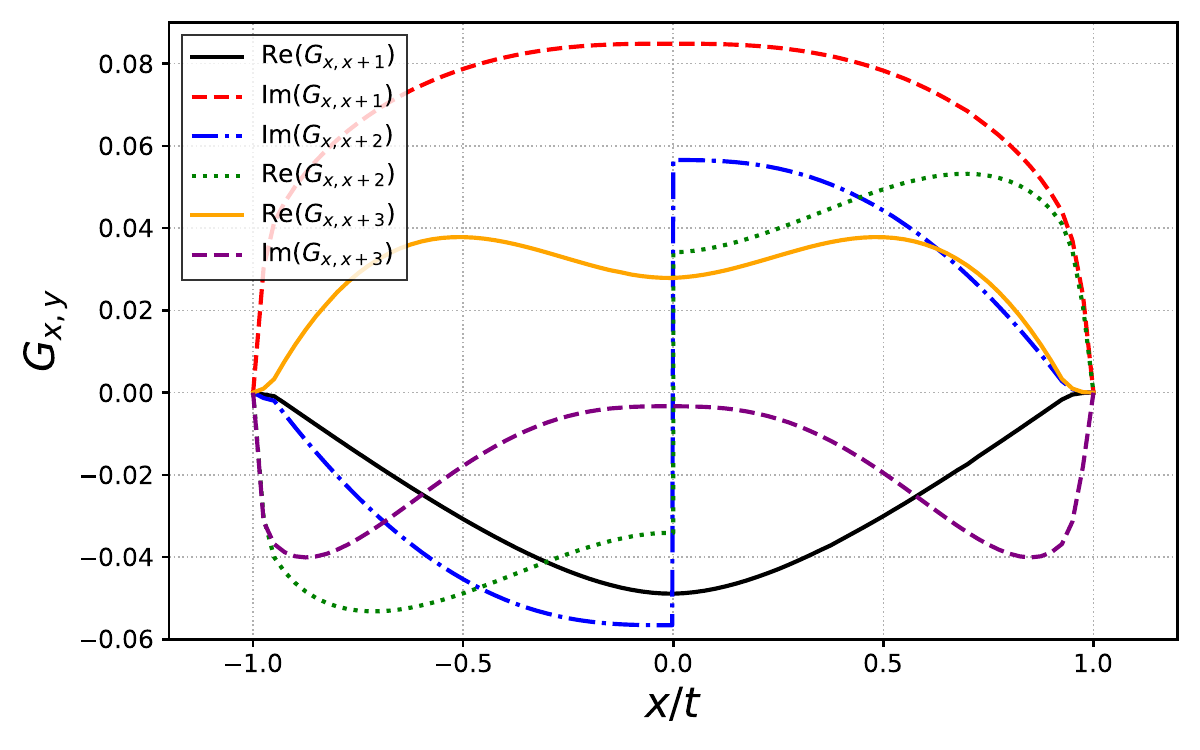}
\caption{ 
 Dynamics of the two-point correlation functions $G_{xy}(t)$ for $y=x+j$, with $j=1,2,3$. 
 We show both the real and imaginary parts of the correlator versus $x/t$. The curves are 
 the analytical predictions  obtained in Section~\ref{sec:hydro} in the hydrodynamic limit 
 $x,t\to\infty$ with fixed $x/t$. We focus on the inhomogeneous XX chain with $h_L=1$ and 
 $h_R=1/3$. The results are for the quench from the initial state~\eqref{eq:dneel-state}, 
 in which the right-part of the chain is prepared in the N\'eel state and the left in the vacuum 
 state.  
}
\label{fig:scaling}
\end{figure} 
%############################################
%
In Fig.~\ref{fig:scaling} we illustrate the theoretical predictions for $G_{xy}$ focusing on the 
off-diagonal elements $G_{x,x+j}$ with $j=1,2,3$. In the figure we show both the real and imaginary parts 
of $G_{xy}$. We focus on the quench from the state~\eqref{eq:dneel-state}. The curve correspond to $G_{xy}^{\scriptscriptstyle[+++]}
+G^{\scriptscriptstyle[+--]}_{xy}$. It is interesting to observe that in the hydrodynamic limit 
$G_{x,x+j}$ exhibits a discontinuity at the origin for even $j$. 

Let us consider the correlations between points $x>0$ and $y<0$ across the interface, which are crucial to 
understand entanglement dynamics. One can check that we have to consider only the contribution with $j=2$ in~\eqref{eq:Sj-compact}, 
because the one with $j=3$ gives vanishing contributions in the 
hydrodynamic limit, and that with $j=1$ is exponentially suppressed in the large $x$ limit. 
The only term that contributes in the hydrodynamic limit is $\propto \rho_+^x\bar\lambda_+^{-y-1}$
This means that the off-diagonal correlator $G_{xy}^{\scriptscriptstyle[++-]}$ is given as 
\begin{multline}
	G_{xy}^{\scriptscriptstyle[++-]}=
	\sum_k\int\frac{ds}{2\pi}\int\frac{ds'}{2\pi} e^{-i(s-s')t}
	\frac{2}{\rho_+(s)-\rho_-(s)}\\  \rho_+^x 
	\left( \rho_+^{-k}(s) + \frac{\rho_+^k(s)(\rho_+(s)-\lambda_+(s))+
			\rho_+^{k+1}(s)(1-\lambda_-(s)\rho_+(s))}{2(h_R-h_L)} 
		\right)\\
		\frac{\bar\rho_+^k(s')\bar \lambda_+^{-y-1}(s')(\bar\lambda_+(s')-\bar\rho_+(s'))}{h_R-h_L}, 
\end{multline}
The term with $\rho_+^{-k}$ in the round bracket does not contribute in the hydrodynamic limit. 
The steps to derive the result are the same as before. Finally, we obtain 
\begin{multline}
\label{eq:Gxy-ppm}
G_{xy}^{\scriptscriptstyle[++-]} = -\frac{1}{2} \int_{0}^{\pi} \frac{dk}{2\pi} \,
\Theta\!\left( -\frac{x}{2} + t v(k) + \frac{(1+y)v(k)}{2v(k')} \right)\\ 
 \exp\!\Big( i(-k+\pi)x + i(1+y)(\pi-k') \Big) 
 \exp\!\left( -\frac{i}{2}(k-k') \right) 
 \frac{\sin\!\left( \frac{k-k'}{2} \right)}{\sin k'}\mathrm{Re}[T(\pi-k)], 
\end{multline}
where $k'$ is determined from the conservation of energy across the 
interface as 
\begin{equation}
	k'=\arccos(h_L-h_R+\cos(k)). 
\end{equation}
Finally, we have to determine the contribution obtained from summing over negative $k$ in~\eqref{eq:ansatz}, 
i.e., $G_{xy}^{\scriptscriptstyle[-+-]}$. The result is obtained from~\eqref{eq:Gxy-ppm} by 
exchanging $x\leftrightarrow y$ and $h_L\leftrightarrow h_R$ as 
\begin{multline}
	\label{eq:Gxy-mpm}
	G_{xy}^{\scriptscriptstyle[-+-]}=
	-\frac{1}{2}\int_0^{\pi}\frac{dk}{2\pi}\Theta\left(-\frac{y}{2}+v(k)t+\frac{(x+1)v(k)}{2v(k')}\right)\\
	\exp\left({i(\pi-k)y+i(\pi-k')x}\right)\exp\left({-\frac{i}{2}(k-k')}\right)
	\frac{\sin\left(\frac{k-k'}{2}\right)}{\sin(k')}\mathrm{Re}[T(k)],
\end{multline}
where $k'=\arccos(h_R-h_L+\cos(k))$. 

To understand better the hydrodynamic interpretation of the results it is convenient to 
consider a mesoscopic cell of size $2\Delta \ell$ centered around point $x$. Here 
$1\ll\Delta\ell\ll L$. The correlation matrix in momentum space between cells is 
\begin{equation}
	G_{x,y}^{\scriptscriptstyle[+++]}=\sum_{x,y} e^{-ik x+iqy}G_{X+x,Y+y}, 
\end{equation}
where $X,Y$ are the positions of the mesoscopic cells and $x,y\in[-\Delta,\Delta]$, and 
$k,q\in[-\pi,\pi]$. We obtain 
\begin{align}
	\label{eq:cell-1}
	& \widetilde G_{XY}^{\scriptscriptstyle[+++]}
	=\frac{\delta(k-q)}{2}\left[\mathbf{1}_{[-\pi,0]}+\mathbf{1}_{[0,\pi]}
-\mathbf{1}_{[0,\pi]}\Theta(-Z+v(\pi-k)t)\mathrm{Re}[T(k)]\right]e^{ik(X-Y)},\\
\label{eq:cell-2}
	& \widetilde G_{XY}^{\scriptscriptstyle[+--]}
	=\frac{\delta(k-q)}{2}\mathbf{1}_{[-\pi,0]}\Theta(Z+v(k-\pi)t)\mathrm{Re}[T(k-\pi)]e^{ik(X-Y)},\\
	\label{eq:cell-3}
	& \widetilde G_{XY}^{\scriptscriptstyle[-++]}
	=\frac{\delta(k-q)}{2}\mathbf{1}_{[0,\pi]}\Theta(-Z+v(-k-\pi)t)\mathrm{Re}[T(k)]e^{ik(X-Y)},\\
	\label{eq:cell-4}
	& \widetilde G_{XY}^{\scriptscriptstyle[---]}
	=\frac{\delta(k-q)}{2}\Big(\mathbf{1}_{[-\pi,0]}+\mathbf{1}_{[0,\pi]}-\mathbf{1}_{[-\pi,0]}\Theta(Z+v(k+\pi)t)\mathrm{Re}[T(k+\pi)]\Big)
	e^{ik(X-Y)}, 
\end{align}
where $\mathbf{1}_{[-\pi,0]}$ and $\mathbf{1}_{[0,\pi]}$ denote the $k$-domain, and $Z=(X+Y)/2$. All the contributions 
are diagonal in momentum space. The phase factor $e^{ik(X-Y)}$ is the same for all the terms. In~\eqref{eq:cell-1} 
the first and the second term correspond to left and right moving quasiparticles, respectively. The last term in~\eqref{eq:cell-1} 
accounts for the fact that the density of right-moving quasiparticles is reduced because some left moving quasiparticles that 
scatter at the origin are transmitted. The term in~\eqref{eq:cell-2} describes the left moving quasiparticles that 
are transmitted to the left. The terms~\eqref{eq:cell-3} and~\eqref{eq:cell-4} are obtained from~\eqref{eq:cell-1} and~\eqref{eq:cell-2} 
by exchanging left and right moving quasiparticles. 
Let us now discuss the case of correlations between cells across the origin, i.e., for $X>0$ and $Y<0$. 
After performing the Fourier transform of~\eqref{eq:Gxy-ppm}, we now obtain that 
the momenta $k,q$ are related as 
\begin{equation}
	q=k'(\pi-k)-\pi=\arccos(h_L-h_R-\cos(k))-\pi. 
\end{equation}
The final result is 
\begin{multline}
	\label{eq:Gxy-across}
\widetilde G_{XY}^{\scriptscriptstyle[++-]}=
-\mathbf{1}_{[0,\pi]}\frac{\delta(q-k'(\pi-k)-\pi)}{2} \Theta\!\left( -\frac{X}{2} + t v(k) + \frac{Y v(k)}{2v(q+\pi)} \right)\\ 
\exp\!\Big(i k X-i q Y \Big) 
\exp\!\left( \frac{i}{2}(q+k) \right) 
\frac{\sin\!\left( \frac{q+k}{2} \right)}{\sin(q)}\mathrm{Re}[T(k)]. 
\end{multline}
Now, the correlator is not diagonal. If one takes limit $h_L=h_R$ the delta function 
becomes $\delta(q+k)$, meaning that cells in different sides of the chain are correlated by entangled 
left and right moving quasiparticles. 
The dynamics of the correlator starting from the homogeneous N\'eel state is obtained as $G_{XY}^{\alpha,\beta}=G_{XY}^{+\alpha\beta}+
G_{XY}^{-\alpha\beta}$, with $\alpha,\beta=\pm$ the signs of $X$ and $Y$. From~\eqref{eq:cell-1} and~\eqref{eq:cell-2} we obtain 
\begin{equation}
	\label{eq:neelGdiag}
	\widetilde G_{XY}^{\scriptscriptstyle[++]}=G_{XY}^{\scriptscriptstyle[--]}
	=\frac{\delta(k-q)}{2}\Big(\mathbf{1}_{[-\pi,0]}+\mathbf{1}_{[0,\pi]}\Big)e^{ik(X-Y)},
\end{equation}
where the superscripts $\pm$ are the signs of $X,Y$. Eq.~\eqref{eq:neelGdiag} reflects that 
the diagonal correlators remain the same as in the N\'eel state because the net flux of quasiparticles 
across the origin is zero. 

%############################################
\subsection{Quasiparticle picture for the von Neumann entropy}
\label{sec:vN}

Finally, we should observe that the Fourier transform of the correlators with $X,Y>0$ allows to obtain 
the dynamics of the von Neumann entropy of any subsystem $A$ within the right-half of the system. However, 
this would require to determine the behavior of the correlators for $X,Y>0$ with $X,Y$ far apart from each other, 
which are subleading in the large $X,Y$ limit. Physically, for the dynamics from the N\'eel state they describe the correlation between the entangled 
pairs that are formed after the quench and propagate in opposite directions. These correlations are present because 
the sum over $k$ in~\eqref{eq:ansatz} gives rise to vanishing denominators as $1-e^{2i B(s,s')}$ (cf.~\eqref{eq:Gxy-pmm}, for 
an example), where the $2$ reflects the two-site translation invariance of the N\'eel state, and $B(s,s')$ is a function. 
The factor $2$ in the exponential implies that there are poles at $B(s,s')=0$ and $B(s,s')=\pi$. In deriving~\eqref{eq:Gxy-pmm}-\eqref{eq:Gxy-mpm} 
we considered only the contribution of the pole at $B(s,s')=0$, which is sufficient to describe the behavior 
of all the local correlators, i.e., for $x\approx y$, and the leading behavior of $G_{xy}$ across the origin (cf.~\eqref{eq:Gxy-ppm} and 
~\eqref{eq:Gxy-mpm}). To extract entanglement properties one would need to determine the contribution of the pole at 
$B(s,s')=\pi$. This would give the Fourier transform of the full correlator between different mesoscopic cells 
at $X,Y$ in the right-half region, which allows to determine the entanglement dynamics by employing the strategy of Ref.~\cite{alba2021unbounded} 
(see also~\cite{caceffo2023entanglement}). Notice that, at least in principle, following Ref.~\cite{caceffo2026fate} it should be possible to 
conjecture the quasiparticle picture from the Fourier transform of the correlator, provided that a quasiparticle picture 
exists (see for instance, Refs.~\cite{parez2026reduced,parez2026smearing} for situations where the quasiparticle picture is not applicable). 

Finally, although our results do not allow us to obtain the quasiparticle picture for the quench from the 
homogeneous N\'eel state, they allow us to prove the results of Ref.~\cite{dipasquale2026entanglement}, i.e., for the 
quench starting from~\eqref{eq:dneel-state}. Precisely, 
one can use Eq.~\eqref{eq:Gxy-pmm} to compute the von Neumann entropy of a subsystem $A$ of length $\ell$ placed 
next to the origin on the negative side. Since the global state of the system is pure, this equals the von Neumann 
entropy of a subsystem of length $\ell$ placed next to the origin on the positive side. One can employ the 
approach of Ref.~\cite{alba2021unbounded} to derive \textit{ab initio} the result of Ref.~\cite{dipasquale2026entanglement} 
from the correlator~\eqref{eq:Gxy-pmm}. On the other hand, following Ref.~\cite{bertini2018entanglement} (see also~\cite{caceffo2026fate}), 
one can suggestively write the reduced density matrix of a region $A=[-\ell,0]$ as 
\begin{equation}
	\label{eq:sc-rho}
	\rho_A(t)=\bigotimes_{X\in A}\bigotimes_k\rho_{X,k}, 
\end{equation}
where the tensor product is over the different mesoscopic cells contained in $A$ and over the momentum $k$. 
In~\eqref{eq:sc-rho} $\rho_{X,k}$ is the density  matrix describing correlations inside the cell at $X$. Since 
the cells are described by Gaussian states, $\rho_{X,k}$ is identified by the correlation matrix, which in this 
case is $G^{\scriptscriptstyle[+--]}_{XX}$ (cf.~\eqref{eq:cell-2}), which if $X+v(k-\pi)t>0$, i.e., if the transmitted 
quasiparticle from the right region is in the cell, describes a mixed state. If the quasiparticle is not in the cell, 
this is in a pure state. Now, to compute the von Neumann entropy one has to take the trace as 
\begin{equation}
	S_A=-\mathrm{Tr}\rho_A\ln(\rho_A), 
\end{equation}
where to perform the trace one has to sum over the positions $X$ of the cells and over the momenta $k$. The sum 
over $X$ gives $\min(v(k)t,\ell))$. Finally, one obtains 
\begin{equation}
	\label{eq:conj}
	S_A(t)=\int_{0}^\pi\frac{dk}{2\pi}\min(v(k)t,\ell) s(\mathrm{Re}(T(k))/2),
\end{equation}
where $s(x)=-x\ln(x)-(1-x)\ln(1-x)$. Eq.~\eqref{eq:conj} coincides with the main result of Ref.~\cite{dipasquale2026entanglement}.

%
%############################################
\section{Numerical benchmarks}
\label{sec:numerics}

%
%############################################
\begin{figure}[t]
\centering
\includegraphics[width=.85\linewidth]{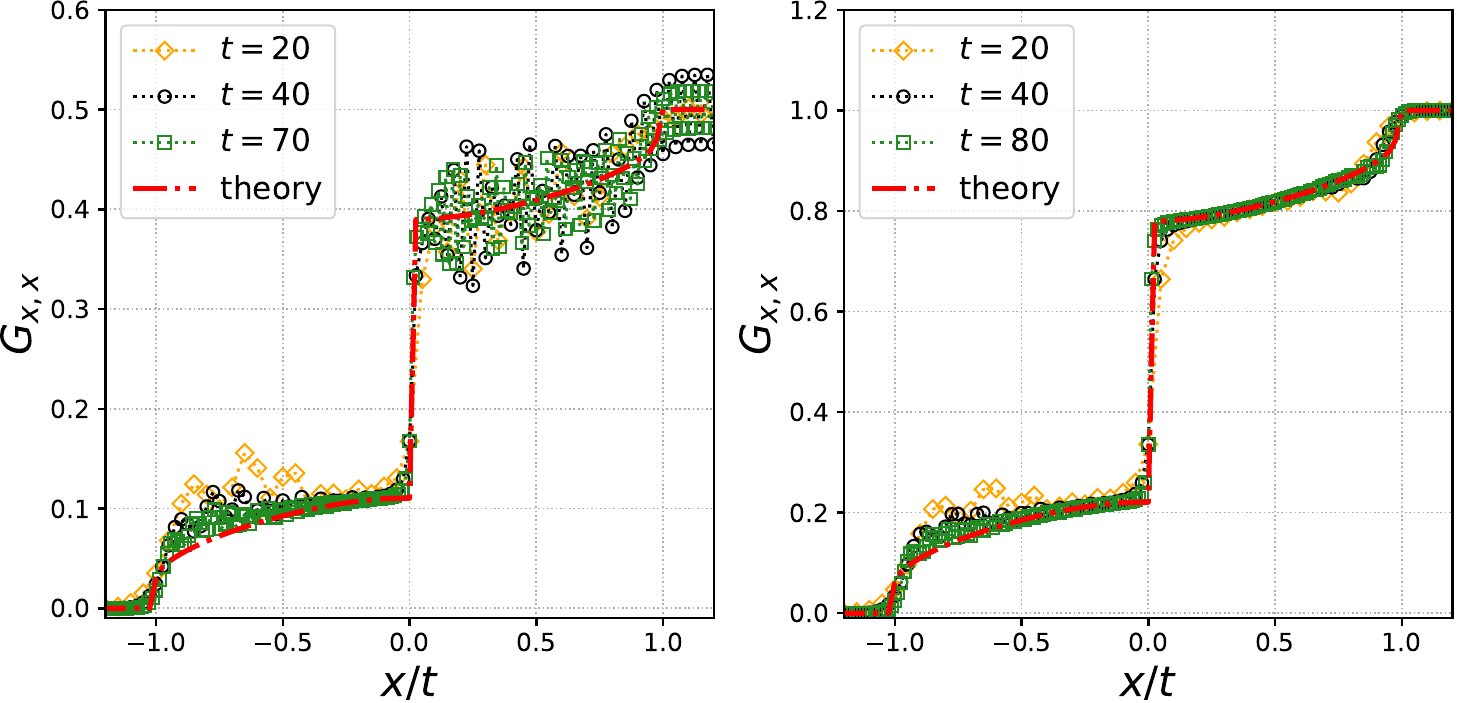}
\caption{ Dynamics of the fermion density after the quenches from the state~\eqref{eq:dneel-state} 
	(left panel) and~\eqref{eq:d-state} (right panel) in the inhomogeneous XX chain with $h_L=1$ and 
	$h_R=1/3$. The symbols are exact numerical data obtained by solving the system~\eqref{eq:one} for 
	several times. The dashed-dotted line is the analytic result in the hydrodynamic limit.  The results 
	are for the XX chain with $h_L=1$ and $h_R=1/3$. 
}
\label{fig:neel-DW-diag}
\end{figure} 
%############################################
%

Let us now discuss numerical benchmarks of the results of Section~\ref{sec:hydro}. 
In Fig.~\ref{fig:neel-DW-diag} we focus on the fermionic density $G_{xx}$. We consider 
the quench from the state~\eqref{eq:dneel-state} and the domain wall state~\eqref{eq:d-state} 
(see left and right panel in the figure, respectively). In the figure we consider the quench 
with the inhomogeneous XX chain with $h_L=1$ and $h_R=1/3$. The symbols in Fig.~\ref{fig:neel-DW-diag} 
are exact numerical data obtained by using~\eqref{eq:one} for several times. The dashed-dotted line 
is the theoretical prediction in the hydrodynamic limit obtained by summing~\eqref{eq:Gxy-ppp} and~\eqref{eq:Gxy-pmm} 
for the quench from~\eqref{eq:dneel-state}. For the quench from~\eqref{eq:d-state} the result is 
twice that for the quench from~\eqref{eq:dneel-state}. The data for the quench with N\'eel state (left panel 
in Fig.~\ref{fig:neel-DW-diag}) show strong deviation from the hydrodynamic limit, which are likely due to the 
fact that $t$ is not large enough to reach the hydrodynamic limit. This is compatible with the results 
of Ref~\cite{alba2022noninteracting} for quenches from the N\'eel state in the presence of dissipative 
impurities. On the other hand, for the quench from the domain wall state~\eqref{eq:d-state} (right-panel in 
Fig.~\ref{fig:neel-DW-diag}) finite-time corrections are smaller and the agreement with the theoretical predictions 
is satisfactory already for $t=80$. 

%
%############################################
\begin{figure}[t]
\centering
\includegraphics[width=.95\linewidth]{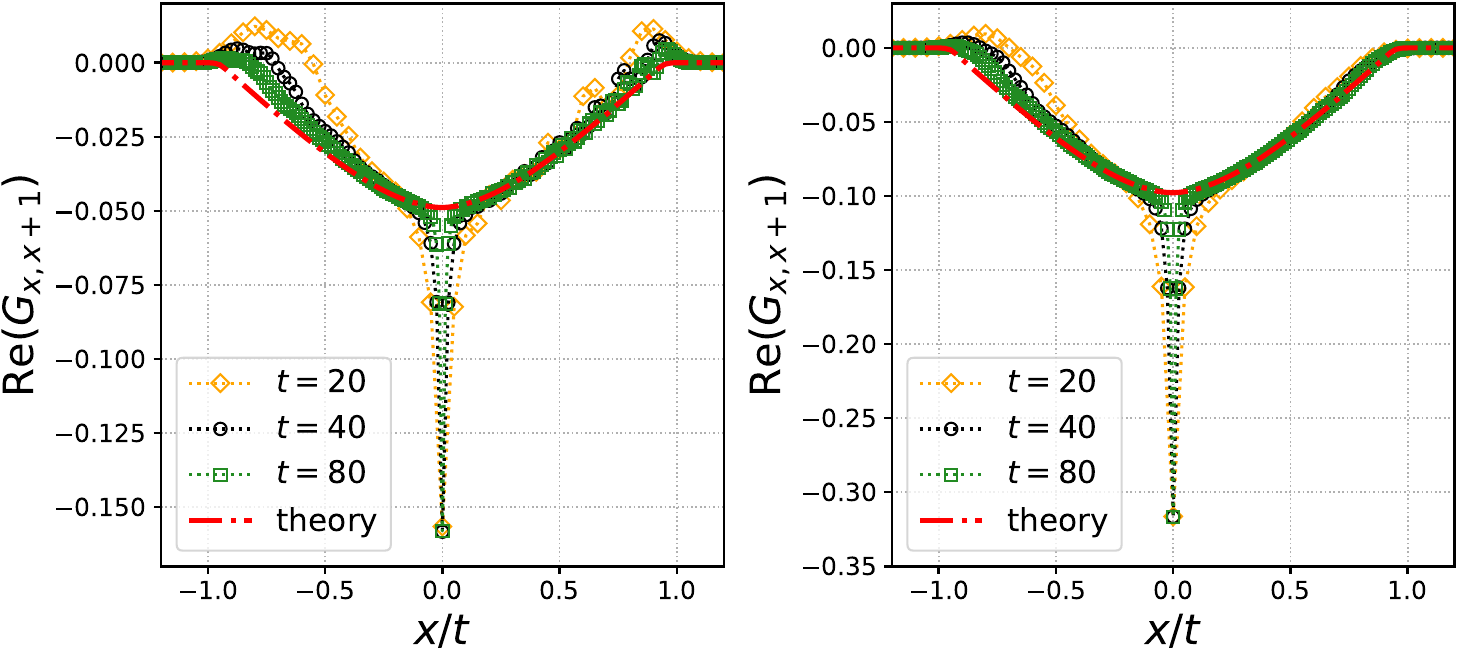}
\caption{ Dynamics of the off-diagonal correlator $\mathrm{Re}(G_{x,x+1})$ in the XX chain with $h_L=1$ 
	and $h_R=1/3$. The left and right panels show results for the quenches from the state~\eqref{eq:dneel-state} and 
	~\eqref{eq:d-state}, respectively. The symbols are exact numerical results, whereas the dashed-dotted line 
	is the analytic result in the hydrodynamic limit. Similar to Fig.~\ref{fig:neel-DW-diag} the behavior at 
	$x/t=0$ is singular. 
}
\label{fig:neel-DW-curr}
\end{figure}
\begin{figure}[t]
\centering
\includegraphics[width=.85\linewidth]{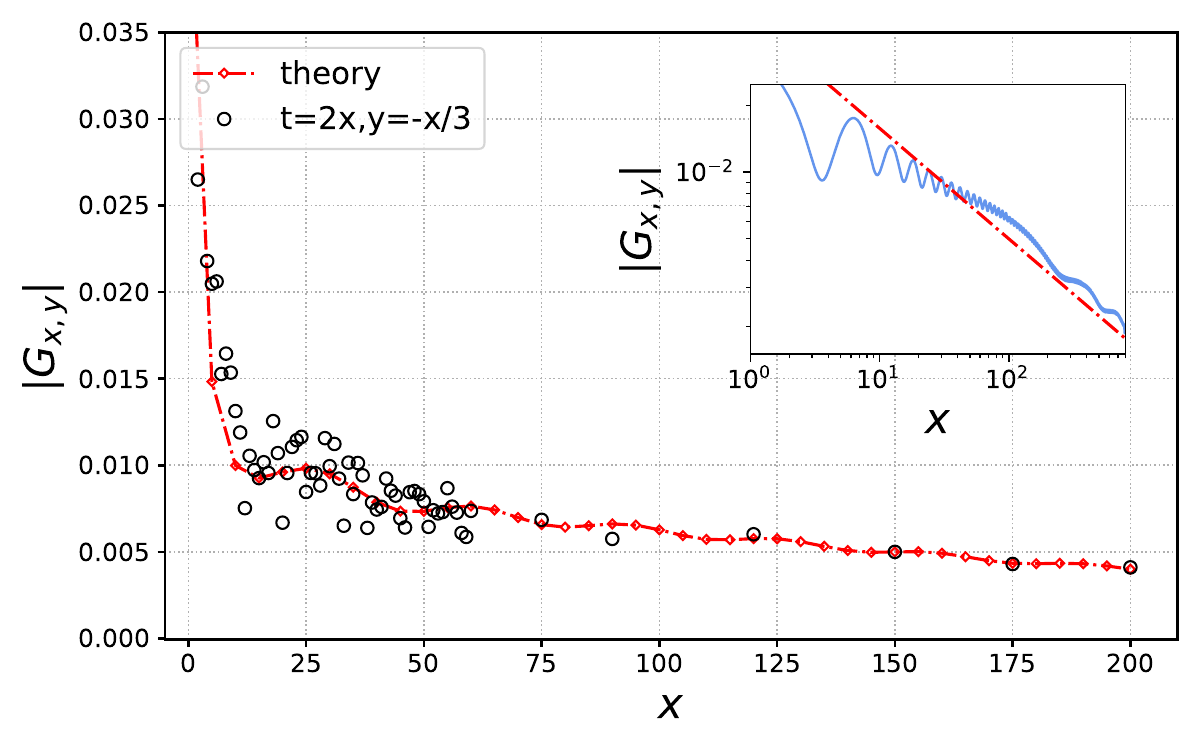}
\caption{ The off-diagonal fermionic correlator $G_{xy}$ after the quench from 
	the state~\eqref{eq:dneel-state} in the XX chain with $h_L=1$ and $h_R=1/3$. 
	We plot $|G_{xy}|$ versus $x$ at fixed $t=2x$ and $y=-x/3$. The circles 
	are exact numerical data. The dashed-dotted line is the result in the hydrodynamic 
	limit~\eqref{eq:Gxy-ppm}. The inset shows the large $x$ behavior of~\eqref{eq:Gxy-ppm} as 
	$G_{xy}^{\scriptscriptstyle[++-]}\sim x^{-1/2}$ (dashed-dotted line). 
}
\label{fig:Gxy-off}
\end{figure}

 
%############################################
%
In Fig.~\ref{fig:neel-DW-curr} we discuss off-diagonal elements of $G_{xy}$. Precisely, we focus on 
$\mathrm{Re}(G_{x,x+1})$. The setup is the same as in Fig.~\ref{fig:neel-DW-diag}. The left and right panels show results 
for the quench from~\eqref{eq:dneel-state} and from~\eqref{eq:d-state}, respectively. For both quenches the numerical data 
exhibit a singularity at the origin $x/t=0$. Again, this is similar to what observed in quenches in the presence of 
dissipative impurities in Ref.~\cite{alba2022noninteracting}. The theoretical prediction is continuous at the origin. 
For both quenches the agreement with the hydrodynamic limit result (dashed-dotted line) is excellent already at 
$t=80$. Finally, let us discuss correlations between points across the interface. In Fig.~\ref{fig:Gxy-off} we focus 
on $G_{xy}$ with $x>0$ and $y<0$. We restrict ourselves to the situation with $t=2x$ and $y=-x/3$, considering the large $x$ 
limit. In Fig.~\ref{fig:Gxy-off} we plot $|G_{xy}|$ versus $x$. The circles are exact numerical data, whereas the triangles 
are the results in the hydrodynamic limit (cf.~\eqref{eq:Gxy-ppm}). The correlator decays in the limit $x\to\infty$. 
Corrections due to the finite $x$ are ``strong'' even for $x\approx 50$, although upon increasing $x$ the data clearly 
approach the analytic result. In the inset we clarify the large $x$ behavior as $x^{-1/2}$ of the correlator.

%
%############################################
%############################################
%

%############################################
\section{Conclusions}
\label{sec:conc}

In this work we provided a complete analytical solution for the dynamics
of the two-point fermionic correlation function $G_{xy}(t)$
in the inhomogeneous tight-binding chain with a step potential
after quantum quenches from product initial states.
The Hamiltonian is quadratic in fermions,
with different magnetic fields $h_L$ and $h_R$ on the two halves of the chain.
The dynamics of $G_{xy}$ is governed by a linear system of equations,
which we solved exactly by introducing the  factorisation~\eqref{eq:ansatz}. 
The dynamics was obtained by reducing the problem
to a Riemann-Hilbert problem on the unit circle,
whose solution yields explicit expressions in terms of elementary functions
after inverse Laplace and Fourier transforms.
Besides exact results for the dynamics of the correlator, we derive the behavior in the 
hydrodynamic limit $x,y,t\to\infty$ with fixed ratios $x/t,y/t$.
We benchmarked our analytic predictions against exact numerical simulations
of the finite-chain dynamics. The agreement is excellent
in the hydrodynamic limit, with finite-size corrections. The numerical data confirm
the validity of the stationary-phase approximation 
and the quasiparticle interpretation underlying our approach.

Several directions for future research emerge from this work.
First, it would be interesting to extend our analysis to other initial states,
such as Fermi seas with arbitrary filling. 
Second, one could apply the same Riemann-Hilbert method
to other quadratic models with inhomogeneities,
such as the Kitaev chain.
Third, the off-diagonal correlators we have obtained
are the natural building blocks for computing
the entanglement entropy and the logarithmic negativity
across the interface. 
Fourth, it would be interesting to investigate
the effect of interactions or dissipation
on the correlation functions, using the framework
of Generalized Hydrodynamics or Lindblad master equations.
Our results are directly relevant to experiments
with ultracold atoms or quantum simulators,
where inhomogeneous potentials can be engineered
and the two-point correlation functions can be measured. 
Finally, it would be interesting to  consider the effects of monitoring
(see, for instance, \cite{bayonapena2026generalized,turkeshi2022enhanced}).
Our work provides a solid foundation for the systematic derivation of the 
quasiparticle picture in inhomogeneous integrable systems.

%############################################
\section*{Acknowledgements}
VA and FR have been supported by the project “Artificially devised many-body quantum dynamics in low dimensions - ManyQLowD” funded by the MIUR Progetti di Ricerca di Rilevante Interesse Nazionale (PRIN) Bando 2022 - grant 2022R35ZBF. This article is based upon work from COST Action ``Many-body Open Quantum Systems'' (QOpen) CA24109 supported by COST (European Cooperation in Science and Technology). VA gratefully acknowledges support from the Simons Center for Geometry and Physics, Stony Brook University, at which some or all of the research for this paper was performed during the Program ``Complexity, information, and tractable simulations of quantum many-body dynamics'' 01/06-02/07 2026.

\bibliography{bibliography.bib}
\nolinenumbers

%\appendix

\end{document}